\documentclass[twocolumn,showpacs,preprintnumbers,amsmath,amssymb, superscriptaddress, prb]{revtex4-1}
\usepackage{amsmath}
\usepackage{graphicx}
\usepackage{dcolumn}
\usepackage{bm}

\usepackage{color}
\definecolor{red}{rgb}{1,0,0}

\definecolor{blue}{rgb}{0,0,1}

\definecolor{green}{rgb}{0,1,0}

\begin{document}
\preprint{APS}

\author {Mariano de Souza}
\email{E-mail: mariano@rc.unesp.br; Present Address: Institute of Semiconductor and Solid State Physics, Johannes Kepler University Linz, 4040 Linz, Austria.}
\affiliation{IGCE, Unesp - Univ Estadual Paulista, Departamento de F\'{i}sica, 
 13506-900, Rio Claro, SP, Brazil}
 \author {Paulo Menegasso}
\affiliation{IGCE, Unesp - Univ Estadual Paulista, Departamento de F\'{i}sica, 
 13506-900, Rio Claro, SP, Brazil}
\author {Ricardo Paupitz}
\affiliation{IGCE, Unesp - Univ Estadual Paulista, Departamento de F\'{i}sica, 
 13506-900, Rio Claro, SP, Brazil}
\author {Antonio Seridonio}
\affiliation{IGCE, Unesp - Univ Estadual Paulista, Departamento de F\'{i}sica, 
 13506-900, Rio Claro, SP, Brazil}
\author {Roberto E. Lagos}
\affiliation{IGCE, Unesp - Univ Estadual Paulista, Departamento de F\'{i}sica, 
 13506-900, Rio Claro, SP, Brazil}

\title{Gr\"uneisen Parameter for Gases}

\vspace{0.7cm}
\begin{abstract}

The Gr\"uneisen ratio ($\Gamma$), i.e.\,the ratio of the linear thermal expansivity to the specific heat at constant pressure, quantifies the degree of  anharmonicity of the potential governing the physical properties of a system. While $\Gamma$ has been intensively explored in solid state physics, very little is known about its behavior for gases. This is most likely due to the difficulties posed to carry out both thermal expansion and specific heat measurements in gases with high accuracy as a function of pressure and temperature. Furthermore, to the best of our knowledge a comprehensive discussion about the peculiarities of the Gr\"uneisen ratio is still lacking in the literature. Here we report on a detailed and comprehensive overview of the Gr\"uneisen ratio. Particular emphasis is placed on the analysis of $\Gamma$ for gases. The main findings of this work are: \emph{i)} for the Van der Waals gas $\Gamma$ depends only on the co-volume $b$ due to interaction effects, it is smaller than that for the ideal gas ($\Gamma$ = 2/3) and diverges upon approaching the critical volume; \emph{ii)} for the Bose-Einstein condensation  of an ideal boson gas, assuming the transition as first-order $\Gamma$ diverges upon approaching a critical volume, similarly to the Van der Waals gas; \emph{iii)} for $^4$He at the superfluid transition $\Gamma$ shows a singular behavior. Our results reveal that $\Gamma$ can be used as an appropriate experimental tool to explore pressure-induced critical points.
\end{abstract}


\maketitle
\vspace{1.0cm}

\date{\today}
\section{Introduction}
It is well known from daily life that upon increasing or
decreasing the temperature of a solid, in general, its volume changes. The formation of ice, namely the phase transition of water from liquid-to-solid, consists a classical example of volume change. It is worth to mention that in the case of water, in particular, a negative thermal expansion is observed upon freezing, thus leading the ice to fluctuate in the surface of the liquid. In general terms, in the case of solids, regarding solely the phononic excitations, the phenomenon of thermal expansion is a direct
consequence of the deviation of the lattice potential from the so-called harmonic approximation, i.e.\,temperature changes
bring terms of the lattice potential $U$ with power higher than two into play, for instance $U(x) = ax^2 - bx^3 - cx^4$, being $a$, $b$ and $c$ positive constants and $x$ the average atomic displacement from the equilibrium position and, as a consequence $<x>$ is not longer zero \cite{Kittel}.
Thus, as an effect of such anharmonic contributions to the lattice potential, once the temperature is varied the solid continuously either shrinks or dilates in order to achieve the optimal volume where its total free energy ($F$) is minimized, namely $\partial F / \partial <x> $ = 0. The situation is quite distinct in the immediate vicinity of a generic phase transition, where besides such anharmonic terms contributing to the lattice potential other sorts of excitations like charge, spins and orbital as well as critical fluctuations itself \cite{2015} can contribute dramatically to the volume change of the solid.
\begin{figure}
  \centering
  \includegraphics[scale=0.650]{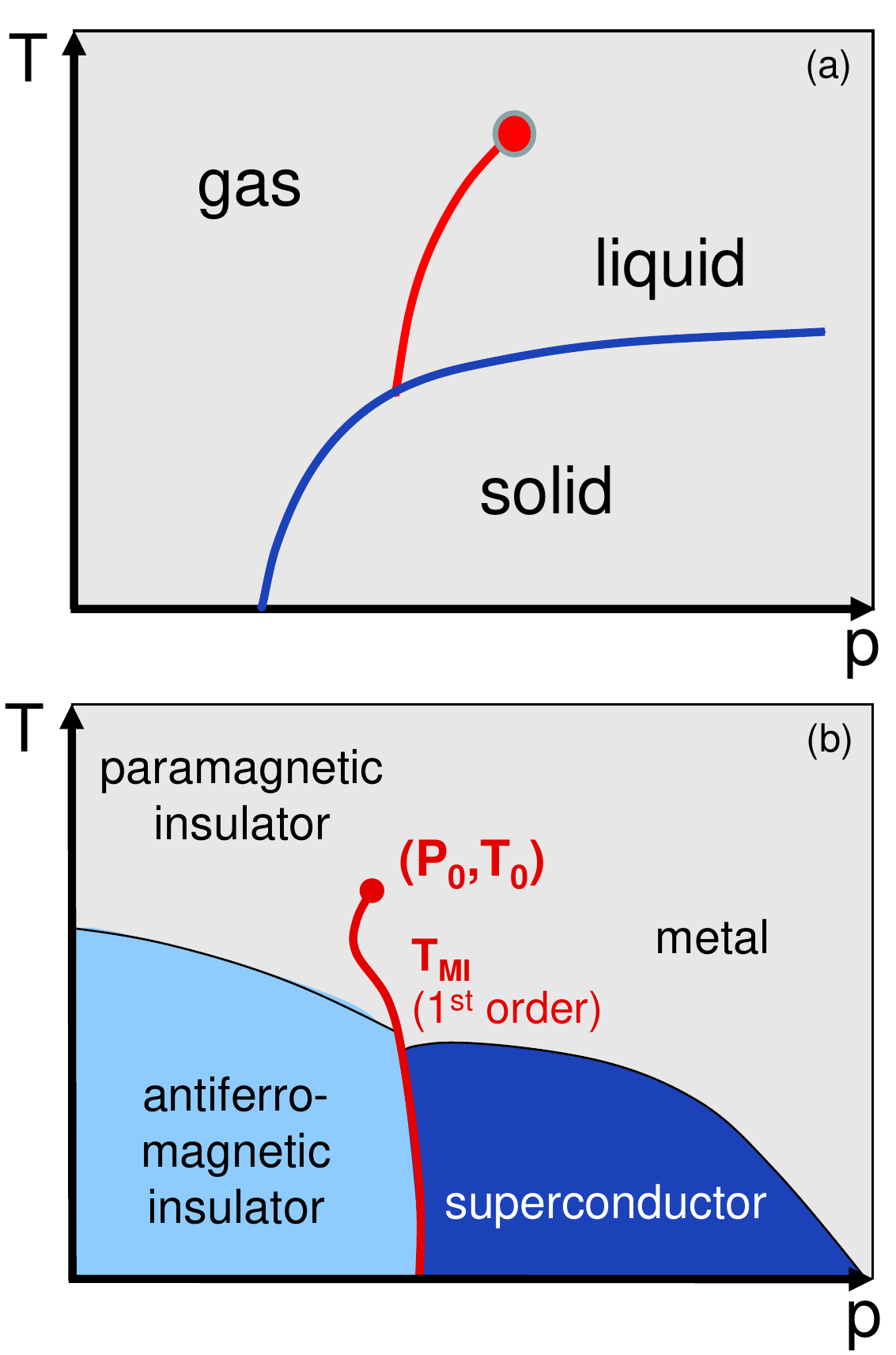}
  \caption{(a) Schematic universal pressure ($p$) \emph{versus} temperature ($T$) phase diagram for a generic substance, being solid, liquid and gaseous phases indicated explicitly. (b) Schematic proposed universal pressure \emph{versus} temperature phase diagram of molecular conductors of the $\kappa$-(BEDT-TTF)$_2$X serie \cite{Mariano, Barto, 2015}. The red solid lines in (a) and (b) represent a first-order phase transition line which separate the paramagnetic/antiferromagnetic insulating from the metallic/superconducting phased and ends in a critical point (indicated by red bullets in (a) and in (b)), indicated by ($P_0$, $T_0$). Details about the behavior of the GR close to the critical-end point are presented in the main text.}\label{Fig-2}
  \end{figure}
From the experimental point of view, such volume changes cannot, in many cases, be detected directly by means of standard x-ray structural data analysis due to the relatively low-resolution intrinsic to this kind of experiments, i.e.\,$\Delta l/l \simeq$ 10$^{-8}$, here $l$ refers to the sample length,  see e.g.\,\cite{Barron}.  In this sense, high-resolution measurements of the thermal expansivity can be considered a quite powerful thermodynamic experimental tool for investigating volume changes associated with phase transitions. Indeed, employing such an experimental technique phase transitions involving distinct types of excitations, i.e.\,lattice, charge, magnetic and orbital degrees of freedom can be precisely detected. For instance, ultra-high resolution expansivity measurements have been recently employed to detect and explore subtle lattice effects accompanying the charge-ordering transition in molecular solids \cite{Mariano08,Foury}, spin-liquid-like lattice instability \cite{SL}, the Mott metal-to-insulator transition \cite{Mariano,Thesis,PRB2012,2015}, magnetic \cite{RSI}, multiferroic \cite{Pregelej} and superconducting transition as well \cite{EPJB}.
Analogously, the bulk properties of a solid can be obtained by means of high-resolution specific heat measurements \cite{2015sh}. Indeed, high-resolution specific heat measurements constitute one of the most powerful experimental tools to access fundamental excitations in solids, like the effective charge carrier mass and the entropy changes associated with a phase transition \cite{Gopal}, just to mention a few examples.
In his seminal paper of 1908 \cite{Grueneisen1908}, E.\,Gr\"uneisen reported on a formal connection between both quantities, i.e.\,thermal expansion and heat capacity, which leads to the birth of the so-called \emph{Gr\"uneisen Ratio}, abbreviated to GR hereafter.  The GR ($\Gamma$) is generally defined as the ratio of the linear thermal expansivity ($\alpha$) to the specific heat at constant pressure ($c_p$), namely:
\begin{equation}
\Gamma = \frac{\alpha}{c_p}.
\label{Definition-GR}
\end{equation}

It is straightforward to write Eq.\,\ref{Definition-GR} in the following way:

\begin{equation}
\Gamma =  V  \frac{\left( \partial P / \partial T \right)_V}{\left( \partial E / \partial T \right)_V} = - \frac{1}{T}
  \frac{\left( \partial S / \partial p \right)_T}{\left( \partial S / \partial T \right)_p},
\label{Definition-GR1}
\end{equation}
where  $V$, $P$, $T$, $E$ and $S$ refer, respectively, to the specimen volume, pressure, temperature, energy and entropy.
For completeness, it is worth recalling that $(\partial E / \partial T)_V$ is the specific heat at constant volume.

\begin{figure}
\center
\includegraphics[scale=0.650]{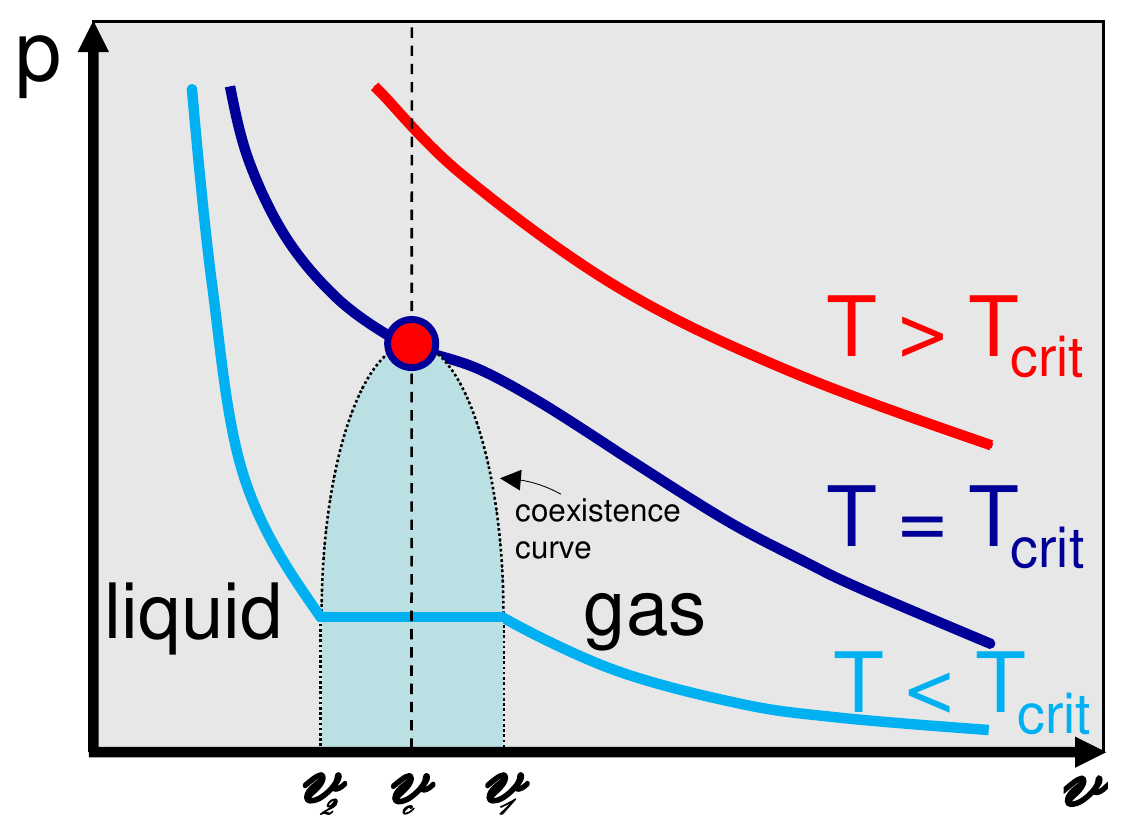}
\caption{Schematic universal pressure \emph{versus} volume phase diagram. The shaded area indicates a phase coexistence region. At $T$ = $T_{crit}$, i.e.\,at the liquid-to-gas phase transition temperature the system achieves a critical volume and the GR diverges. For details, see discussion in the main text.}\label{Fig-3}
\end{figure}

More generally, the GR is roughly constant when a single energy scale $E^s$, for instance the exchange coupling constant in a magnetic system or the electrical polarization in a conventional ferroelectric, governs the physical properties of the system of interest. In such cases, for the sake of completeness, it is worth mentioning that the entropy $S \propto f(T/E^s)$ and the GR are given by:
\begin{equation}
\Gamma = \Bigl(\frac{1}{V_m E^s}\Bigl)\frac{\partial E^s}{\partial p},
\label{Definition-GR-Entropy}
\end{equation}
see e.g.\,\cite{Zhu} and references cited therein for details. However, from the definition of the GR (cf.\,Eqs.\ref{Definition-GR} and \ref{Definition-GR1}) one can directly infer that for \emph{any} pressure-induced critical point the GR should diverge. This is a direct consequence of the entropy accumulation in the immediate vicinity of a critical point.  Indeed, it is nowadays well established that the GR shows a singular behavior upon approaching a quantum critical point \cite{Kuechler, Zhu} and close to a finite-temperature critical end-point \cite{Mariano, Barto, 2015} as well (see Figs.\,\ref{Fig-2} and \ref{Fig-3}). Hence, the GR can considered the \emph{smoking gun} for exploring pressure-induced critical points (see Figs.\,\ref{Fig-2} a), b) and Fig.\,\ref{Fig-3}), no matter the nature of the phase transition.  While the GR has been largely explored in solid state physics, since measurements of the specific heat and thermal expansion measurements are quite accessible, very little is known about its behavior for gases. This is most likely due to the difficulties posed to carry out both thermal expansion and specific heat measurements in gases with high accuracy as a function of pressure and temperature. To the best of our knowledge, the only report about the GR for gases is found in Ref.\,\cite{Gas}. Also in classical textbooks discussions on the GR are quite limited, see e.g.\,Refs.\,\cite{Landau, ashcroft}. Interestingly enough, recently the GR was identified as the scaling exponent $\gamma$  in supercooled liquids \cite{JPCM}. As discussed by the authors of Ref.\,\cite{JPCM} the scaling exponent $\gamma$ quantifies the slope of the interatomic potential and its relative contribution of temperature and volume. Hence, as pointed by the authors in Ref.\,\cite{JPCM} a direct connection between $\gamma$ and the GR exists, characterizing thus both quantities as a \emph{gauge} for the level of the anharmonicity of the potential. 
In this contribution we introduce the fundamental physical concepts associated with the GR in a didactic way and report on a systematic analysis of the GR for gases. We show that for the Van der Waals gas the GR depends only on the co-volume $b$ due to interaction effects, it is smaller than that for the ideal gas (GR = 2/3) and diverges upon approaching the critical volume. For the Bose-Einstein condensation of an ideal boson gas, assuming the transition as first-order the GR diverges upon approaching a critical volume, similarly to the Van der Waals gas. Furthermore, for $^4$He at the superfluid transition the GR shows a singular behavior at the transition.
Hence, we show, for the first time, the particularities of the GR for various gaseous systems. The paper is divided as follows: after this brief introduction, which comprises the first section, we discuss the most relevant textbooks thermodynamic quantities of interest for this work followed by an analysis of the GR for well-known systems like the ideal, Van der Waals and ultra-relativistic gases; $^4$He and the Bose-Einstein condensation (BEC). 
Our aim is to give an overview of the GR and its relevance in the exploration and understanding of the above-mentioned systems in a comprehensive fashion.

\section{Thermodynamic Quantities}
The Gr\"uneisen-Relation, discussed in the introduction of this work, was originally \cite{Grueneisen1908} presented as follows:
\begin{equation}
\beta(T) = \Gamma_{eff} \cdot \frac{\kappa_T}{V_{mol}} \cdot c_V(T),
\label{Grueneisen}
\end{equation}
where $\beta$($T$) and $V_{mol}$ refers, respectively, to the volumetric thermal expansion coefficient and molar volume; $\Gamma_{eff}$ is the so-called effective Gr\"uneisen parameter, $\kappa_T$ stands for the isothermal compressibility. It is worth mentioning that the GR (Eq.\,\ref{Grueneisen}) holds true for solid, liquid and gases. Although this work is focused on the properties of the GR for gases, for the sake of completeness, in the following we  discuss briefly the lattice Gr\"uneisen parameter, obviously present only in solid-state.  In the frame of the Debye-model, the lattice (or phononic)
Gr\"uneisen parameter $\Gamma_{pho}$ reads:
\begin{equation}
\Gamma_{pho} = -\frac{d\ln \Theta_D}{d\ln V},
\label{LatticeGrueneisenParameter}
\end{equation}
where $\Theta_D$ stands for the Debye temperature of the solid. A simple analysis of
Eq.\,\ref{LatticeGrueneisenParameter} reveals that the bigger the lattice
Gr\"uneisen parameter, the higher the volume dependence of the
vibration modes of the lattice. Strictly speaking, as above-mentioned, the lattice
Gr\"uneisen parameter is a measure of the volume dependence of the
anharmonicity of the lattice vibrations, which in turn is
responsible for the lattice contribution to the thermal expansion in
a solid. In general terms, the vibrational free energy, entropy, specific heat and
thermal expansion originate from sums of contributions $f_i$, $s_i$,
$c_i$ and $\alpha_i$ from independent vibration modes of frequency
$\omega_i(V)$, respectively. Hence, the so-called mode Gr\"uneisen parameter
is defined in the following way:
\begin{equation}
\Gamma_{i} = -\frac{d\ln \omega_i}{d\ln V}\label{Grueneisen modes}.
\end{equation}
Thus, according to Eq.\,\ref{Grueneisen modes}, vibration modes
whose frequency, $\omega_i$, decreases or \emph{softens} as the
volume of the solid decreases will result in a negative Gr\"uneisen
parameter and, from Eq.\,\ref{Grueneisen}, such vibration modes, in particular, will be
responsible for a negative contribution to the overall thermal
expansion of the material \cite{NTE}. As a matter of fact, the so-called mode Gr\"uneisen parameter can be related to the interatomic/intermolecular potential via the elastic force constant, namely $k \equiv d^2U/dr^2$, which in the frame of the harmonic approximation is proportional to $\omega^2$. More generally, in addition to the ordinary phononic background contribution to
the thermal expansion of a certain solid material, other contributions, whose
origin might be electronic or magnetic, have to be taken into
account in an accurate estimate of the GR. This is the case especially at low temperatures, where such
contributions may dominate the thermodynamic properties
\cite{Barron}. Hence, in such cases the Gr\"uneisen usually is written in the following way:
\begin{equation}
\Gamma_{eff} = \Gamma_{ph} +  \Gamma_{el} +
\Gamma_{mag} =  \frac{V_{mol}}{\kappa_T}\Bigl(\frac{\beta_{ph}}{c_{ph}} + \frac{\beta_{el}}{c_{el}} + \frac{\beta_{mag}}{c_{mag}}\Bigl),\label{Grueneisen electronic and magnetic}
\end{equation}
where $\beta_{ph}$ ($c_{ph}$), $\beta_{el}$ ($c_{el}$) and
$\beta_{mag}$ ($c_{mag}$) refer to the phononic, electronic and
magnetic contributions to thermal expansivity $\beta$ (specific heat $c$), respectively, while
$\Gamma_{ph}$, $\Gamma_{el}$ and $\Gamma_{mag}$ are the respective
Gr\"uneisen parameters.  Strictly speaking, when various contributions to the GR have to be taken into account the GR reads \cite{Duane}:
\begin{equation}
\Gamma = \frac{V \sum\limits_i c_i \Gamma_i}{\kappa_T \sum\limits_i \Gamma_i},
\label{Duane}
\end{equation}
where $i$ refers to the $i^{th}$ contribution to $\Gamma$.

At this point, it is useful to discuss the GR in quantitative terms.
For the critical temperature of the ordinary metal-to-superconducting phase
transition in the molecular conductor $\kappa$-(BEDT-TTF)$_2$Cu(NCS)$_2$
 $\Gamma \approx $ 40 was obtained \cite{Jens2000-9c}. The latter is roughly by a factor of twenty bigger than those obtained for
ordinary superconducting materials such as Pb, with a $\Gamma$ = 2.4
\cite{Gladstone1969-9c} or even bigger than those measured for the
well-known high-critical temperature cuprate superconductor YBa$_2$Cu$_3$O$_7$ with $\Gamma$ = (0.36 $\sim$ 0.6)
\cite{Meingast1991-9c}. Such a simple comparison reveals the high-sensitivity of the superconducting transition temperature to application of external pressure in the above-mentioned molecular metals, as discussed in quite details in Ref.\,\cite{Lang2004-9c}. Regarding the GR close to $P$-induced critical points, a huge value of $\Gamma \simeq$ 100 was deduced experimentally in the immediate vicinity of a quantum critical point in heavy fermions  \cite{Kuechler}, close to the critical end-point of the metal-to-insulator Mott transition \cite{Mariano,2015} and at the $\gamma \leftrightarrows \alpha$ structural phase transition of Cerium \cite{Cerium}.
For the sake of completeness, analogously, if the tuning parameter of the phase transition is an external magnetic field ($H$), the magnetic Gr\"uneisen parameter,  frequently also called magnetocaloric effect,  reads \cite{Zhu}:
\begin{equation}
  \label{eq:Grueneisenmag1}
  \Gamma_{H} = - \frac{(\partial M/\partial T)_H}{c_H} =  - \frac{1}{T} \frac{(\partial S/\partial H)_T}{(\partial S/\partial T)_H},
\end{equation}
where $M$ refers to the magnetization and $c_H$ to the specific heat under constant external magnetic field. Furthemore, it is worth mentioning that is well documented in the literature that for a magnetic  field-induced quantum critical point, due to the enhancement of the entropy in the critical region a sign change of $\Gamma_{H}$ is expected \cite{Garst}. The latter consists one of the fingerprints of a magnetic field-induced quantum critical point.

\section{Applications and Discussion}
Below we discuss the GR ratio for various gases and make a detailed discussion for each particular case of interest.

\subsection{The Ideal Gas}
We start with the simplest example. For an ideal gas, well known from textbooks, the equation of state
is given by:
\begin{equation}
pV = nRT,
\label{EOS-Gas-ideal}
\end{equation}
and the (internal) energy in turn, reads:
\begin{equation}
E = \frac{3}{2}nRT,
\label{E-Gas-ideal}
\end{equation}
where $n$ and $R$ refers to the number of moles and the universal gas constant, respectively. Employing Eqs.\,\ref{EOS-Gas-ideal}, \ref{E-Gas-ideal} and making use of the definition for the GR discussed in the introduction of this work (Eqs.\,\ref{Definition-GR} and \ref{Definition-GR1}), one obtains $\Gamma$ = 2/3. Thus, for the ideal gas $\Gamma$ is a constant. Analogously, the same result, i.e.\,$\Gamma$ = 2/3, is obtained for the free electron gas at zero temperature, where $E$ = $3/2 VP$. This result can be easily understood by considering the physical definition of the GR, which quantifies the degree of the anharmonicity of the potential. Since for both ideal and free electron gas the intermolecular interaction is not taken into account and thus no phase transition takes place, a constant value for the GR is expected in these cases. With respect to the value 2/3, it is a direct consequence of the number of degrees of freedom in the system. To make this clear, for the ideal gas, remembering that $C_p - C_V = R$, the GR can be written in the following way:

\begin{equation}
\Gamma = \frac{C_p}{C_V}-1.
\label{Gamma_Gas-ideal}
\end{equation}

Now, considering the relation $C_P/C_V = (f+2)/f$, being $f$ the number of degrees of freedom (for the ideal gas $f$ = 3; namely three translational degrees of freedom), from Eq.\,\ref{Gamma_Gas-ideal}, $\Gamma$ = 2/3 is obtained for the ideal gas. For the sake of completeness, below we derive the GR for the ideal gas employing the partition function:

 \begin{equation}
 Z = \left[V\left(\frac{2\pi mk_BT}{h^2}\right)^\frac{3}{2}\right]^N.
 \end{equation}\label{PF_GI}
 Making use of the well-known relations, namely:
 \begin{equation}
 p = \frac{1}{\beta^*} \Bigl(\frac{\partial ln Z}{\partial V}\Bigl)_T,
 \end{equation}
and
 \begin{equation}
 E = - \frac{\partial ln Z}{\partial \beta^*},
 \end{equation}
being $\beta^* = 1/k_BT$ \footnote{Since $\beta$ was used to label the volumetric thermal expansion coefficient, in order to avoid confusion we make use of $\beta^*$ for labelling $1/k_BT$.}, the GR (cf.\,Eq.\,\ref{Definition-GR}) can thus be written as follows:
 \begin{equation}
 \Gamma = -V \frac{\frac{\partial}{\partial T}\left[ \frac{1}{\beta^*}    \frac{\partial}{\partial V}(\ln{Z})       \right]       }{\frac{\partial}{\partial T}\left[-\frac{\partial}{\partial\beta^*}(\ln{Z})\right]        },\label{Z}
 \end{equation}
making the inner derivatives, one obtains:
 \begin{equation}
 \Gamma = -V\frac{\frac{\partial}{\partial T}\left[\frac{1}{\beta^*}\frac{N}{V}\right]                 }{\frac{\partial}{\partial T}\left[-\frac{3N}{2\beta^*}\right]} = \frac{V\left[\frac{Nk_B}{V}\right]}{\frac{3}{2}k_BN},
 \end{equation}
we then achieve the same value obtained for the GR, namely:
\begin{equation}
\Gamma = \frac{2}{3},
\end{equation}
 employing the equation of state as discussed above.

\subsection{The Van der Waals Gas}
In the case of the Van der Waals gas, the equation of state reads:
\begin{equation}
\Bigl(P + \frac{n^2a}{V^2}\Bigl)\Bigl(V - nb \Bigl) = nRT \label{vdwg},
\end{equation}
where $a$ and $b$ stand for the cohesion pressure and co-volume, respectively; while $n$ refers to the number of particles of the gas investigated.
The Van der Waals gas energy is given by:
\begin{equation}
  \label{evdwg}
  U = U_{ideal\,gas} - \frac{n^2a}{V}.
\end{equation}

Making the derivatives of Eqs.\,\ref{vdwg} and \ref{evdwg} the GR for the Van der Waals gas is obtained:
\begin{equation}
  \label{vdW-Gru}
  \Gamma = \frac{2}{3} \cdot \frac{V}{(V - nb)}.
\end{equation}

Interestingly enough, as can be directly inferred from Eq.\,\ref{vdW-Gru}, the GR depends solely on the co-volume. This
means that the cohesion pressure is not relevant and the GR for the
Van der Waals is governed only by the radius of the gas particles.
Yet, from Eq.\,\ref{vdW-Gru}, one can deduce that measuring the
Gr\"uneisen parameter of a real gas, the co-volume $b$ of the investigated gas can be
immediately obtained. Furthermore, upon pressurizing the gas, $b$
increases and, in particular, when $b$ achieves the gas cloud critical
volume  \cite{Sommerfeld} $V_c$ = $nb$ the GR diverges, since the denominator of Eq.\,\ref{vdW-Gru} tends to zero when $V \rightarrow V_c$. As we have done for the ideal gas, for the sake of completeness, below we derive the GR for the van der Waal gas employing the partition function. The latter, for the van der Waals gas, reads:
 \begin{equation}
\ln{Z} = -N\ln{N}+N+N\ln\left[{\frac{(V-Nb)}{\left(\frac{h^2}{2\pi mk_BT} \right)^{\frac{3}{2}}}}\right] + \frac{N^2a}{Vk_BT}
 \end{equation}
and can be rewritten in the following way:
 \begin{equation}
 \ln{Z} = N[1-\ln N+\ln(V-Nb)]-\nonumber
\end{equation}
\begin{equation}
-N\left\{\frac{3}{2}\left[\ln{h^2}-\ln{(2\pi mk_BT)}\right]\right\}
 + \dfrac{N^2a}{Vk_BT}\label{Wando}
 \end{equation}

Hence, making use of Eqs.\,\ref{Z} and \ref{Wando} the GR can be easily calculated:
 \begin{equation}
 \Gamma = -V \frac{\frac{\partial}{\partial T}\left\{\frac{1}{\beta^*}\left[ \frac{N}{(V-Nb)} - \frac{N^2a}{k_BTV^2}    \right]\right\}             }{\frac{\partial}{\partial T}\left[-\frac{3}{2}Nk_BT+\frac{N^2a}{V}\right]},
 \end{equation}
which simplified, reads:
\begin{equation}
 \Gamma = -V\frac{\frac{\partial}{\partial T}\left\{\frac{k_BTN}{(V-Nb)} - \frac{N^2a}{V^2}             \right\}}{-\frac{3}{2}Nk_B},
\end{equation}
thus, the GR for the van der Waals gas is obtained, namely:
 \begin{equation}
 \Gamma =  \frac{2}{3} \cdot \frac{V}{(V-Nb)},
 \end{equation}
which is identical to Eq.\,\ref{vdW-Gru}, deduced starting from the equation of state.

Still considering the Van der Waals gas, it is useful to analyze the behavior of the GR for the Joule-Thomson effect. For the latter, the temperature gradient is given by:
\begin{equation}
  \label{Joule-Thomsom-vdW}
  \Delta T = -\frac{n^2 a}{c_VV}.
\end{equation}

Combining Eqs.\,\ref{vdW-Gru} and \ref{Joule-Thomsom-vdW}, the GR ($\Gamma_{JT}$) for the Joule-Thomson effect can be obtained, namely:

\begin{equation}
  \label{Joule-Thomsom-Gru}
  \Gamma_{JT} = \frac{2}{3}\frac{1}{\bigl(1+\frac{b c_V \Delta T}{n a}\bigl)}.
\end{equation}

Interestingly enough, as can be deduced immediately  from Eq.\,\ref{Joule-Thomsom-Gru}, in the Joule-Thomson effect the van der Waals GR assumes a value closer to those obtained for the ideal gas, i.e.\,$\Gamma$ = 2/3, when the second term in the parenthesis of the denominator tends to zero.

\subsection{Bose-Einstein Condensation}
The Bose-Einstein gas is of particular interest because, in contrast to the  Fermi-Dirac gas, it can undergo a phase transition thought the particles do not interact with each other. As well-known from the literature, the Bose-Einstein condensation is a direct consequence of how the particles occupancy takes place in the case of bosons. To make this clear, let us recall the average number $\langle N \rangle$ of particles in an allowed state $l$:
\begin{equation}
\label{Particlenumbers-BEC}
\langle N \rangle = \sum_{l} \Bigl(\frac{1}{e^{\beta^*(\epsilon_l - \mu)}-1}\Bigl) = \Bigl(\frac{z}{e^{\beta^* \epsilon_l}-z} \Bigl) = \sum_l \langle n_l \rangle,
\end{equation}
where $z = e^{\beta^* \mu}$ is an effective pressure, the so-called fugacity. Since 1 $\leq e^{\beta^* \epsilon_l} \leq \infty$, i.e.\,all states are accessible in this range, to keep  $\langle N \rangle > 0$,  there is a constrain to the fugacity, namely 0 $\leq z \leq$ 1. In addition, a simple analysis of Eq.\,\ref{Particlenumbers-BEC} reveals that $\mu$ should be either zero or negative. Physically, as a consequence of such constrains, one can say that it is \emph{easy} to tot up additional particles to the gas. A particular situation is found when $l$ = 0, i.e.\,$\epsilon_0$ = 0. In such a situation lim$_{z \rightarrow 0} \langle n_l \rangle \rightarrow \infty$ and a phase transition into a Bose-Einstein condensation takes place \cite{Ketterle}. Here we are interested in estimating the GR 
for the condensate. In the case of the Bose-Einstein condensation of an ideal boson gas, the specific heat
for $T \geq T_c$ is given by\cite{Wang}:

\begin{equation}
\label{Specificheat-BEC-Tacima}
c_v =  1.496 N k_B + 0.341 N k_B \Bigl(\frac{T_c}{T}\Bigl)^{1.5} + 0.089 N k_B \Bigl(\frac{T_c}{T}\Bigl)^{3},
\end{equation}
and, according with Ref.\,\cite{Pathria} for $T > T_c$

\begin{equation}\label{Bosinho}
\frac{\partial P}{\partial T} =  \frac{\partial \Bigl[\frac{N}{V} k_B T \frac{g_{5/2}(z)}{g_{3/2}(z)}\Bigl]}{\partial T},
\end{equation}
where $g_{5/2}(z)$ and $g_{3/2}(z)$ are the so-called Bose functions. 

 At this point, it is worth mentioning that only those particles which do not take part in the condensate contribute to the energy of the Bose gas below $T_c$ \cite{Fetter}. Thus, for $T < T_c$, the specific heat of those particles is given by\cite{Wang}:

\begin{equation}
\label{Specificheat-BEC-Tabaixo}
c_v =  1.926  N k_B \Bigl(\frac{T}{T_c}\Bigl)^{1.5};
\end{equation}
and\cite{Fetter}

\begin{equation}
\label{dPdT-BEC-Tabaixo}
\frac{\partial P}{\partial T} = \frac{\partial [0.0851 m^{1.5}  (k_B T)^{2.5} \hbar^{-3} g]}{\partial T},
\end{equation}
where $m$ refers to the mass of the gas particle, $g$ is the degeneracy, $\hbar = h/2\pi$ being $h$ the Planck constant.

Employing Eqs.\,\ref{Specificheat-BEC-Tacima}, \ref{Bosinho}, \ref{Specificheat-BEC-Tabaixo} 
and \ref{dPdT-BEC-Tabaixo},
the GR of the condensate can be estimated.

It turns out that employing such expressions both below and above the Bose-Einstein condensation temperature GR = 2/3, as in the case of the ideal gas. An alternative equation of state was proposed in Ref.\,\cite{Deeney} considering that in the ground state the system holds the kinetic energy associated with the gas cloud right at the condensation temperature. The proposed equation of state below the condensation temperature reads \cite{Deeney}:
\begin{equation}
\label{Deeney}
PV = cVT^{5/2}+0.513\Bigl[1-(T/T_c)^{3/2}\Bigl]Nk_bT_c,
\end{equation}
where $c = g_{5/2}(1)k^{5/2}(m/2\pi\hbar^2)^{3/2}$. Thus, making use of Eqs.\,\ref{Specificheat-BEC-Tabaixo} and \ref{Deeney} the GR follows a linear dependence with $T$ below $T_c$.
In  what follows, we discuss the BEC in analogy with a liquid-to-gas phase phase transition. The critical volume ($v^*$) is given by \cite{Kardar}:
\begin{equation}
\label{Kardar}
v^* = \frac{1}{n^*} = \frac{\lambda^3}{g\xi_{3/2}} = \frac{1}{g}\Bigl[\frac{h}{(2 \pi m k_B T)^{1/2}}\Bigl]^3;
\end{equation}

Considering the transition as being of first order, the Clausius-Clayperon equation can be used:
\begin{equation}
\frac{dT}{dP} =  \frac{\Delta v}{\Delta s} = \frac{T_c (v^* - v_0)}{L}
\end{equation}
where $\Delta v$, $\Delta s$ and $L$, refer to the volume change, entropy change and latent heat, respectively.

The GR reads:
\begin{equation}
\label{GR}
\Gamma = \frac{L v c_v}{T_c (v^* - v_0)}
\end{equation}

From Eq.\,\ref{GR}, when $v_0 \rightarrow v^* \Rightarrow \Gamma \rightarrow \infty$. A similar result from that obtained for the Van der Waals gas.

From the experimental point of view, immense efforts have been put by different groups to measure the thermodynamic properties of a Bose--Einstein condensate. In this regard, we refer to measurements of the specific heat of a weakly interacting gas \cite{Bagnato} and the influence of pressure on a Bose--Einstein condensate in two dimensions \cite{Cho}.

\subsection{Gr\"uneisen Parameter in the Vicinity of the $^4$He $\lambda$-Phase Transition}
Due to its exotic behavior, the theoretical and experimental investigations of the $^4$He isotope physical properties have been a topic of great interest since the middle of the 20$^{th}$ century until nowadays.  Seminal works include the reports by Fairbank and Kellers on the $\lambda$-like (superfluid) transition \cite{Fair} as well as the detailed report by McCarty \cite{NIST}, who investigated the helium physical properties in a broad range of temperature and pressure.
Furthermore, it is worth mentioning experiments of dielectric constant \cite{Niemela}, thermal expansion \cite{Niemela} and specific heat \cite{Hill, Kramers, Roach}, whose results will serve as the basis for the estimate of the GR to be discussed in the following.
Here, for the estimate of the GR under pressure the results of Graywall and Ahlers \cite{greywall} were employed.
\begin{figure}[h]
\centering
\includegraphics[scale=0.25]{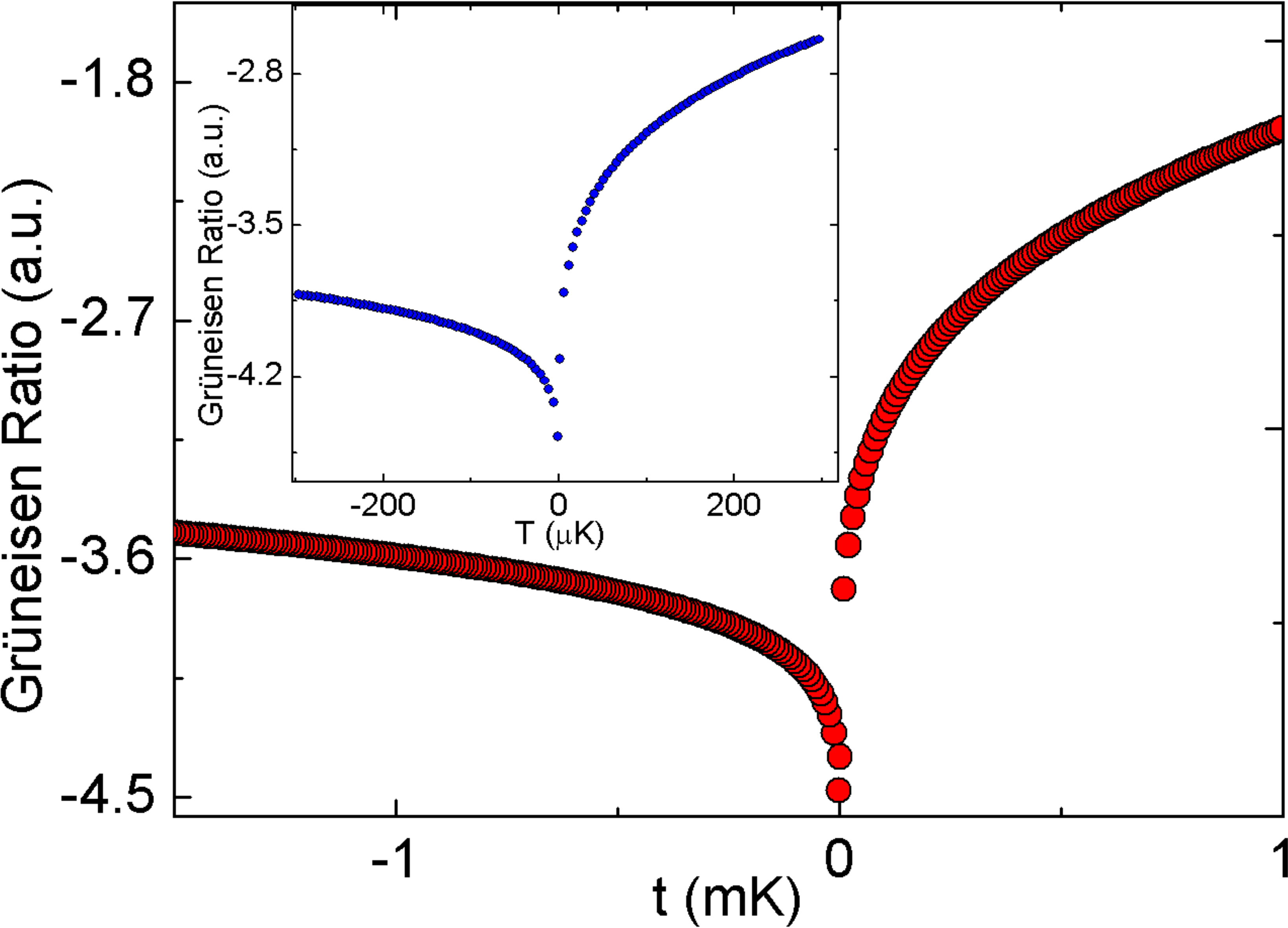}
\caption{\footnotesize Gr\"uneisen ratio as a function of the reduced  temperature $t$  for $^4$He in the immediate vicinity of the normal liquid-to-superfluid phase transition, namely for $\sim|$$t$$|$ $\leq$ 1\,mK.
Inset: Blowup of the data shown in the main panel, being the Gr\"uneisen parameter shown  in the range $|$$t$$|$ $\leq$ 1\,$\mu$K.
}
\label{Gruneisen}
\end{figure}
The expression for the specific heat in the immediate vicinity of the $\lambda$-like transition, reported in the literature by Fairbank and Kellers \cite{Fair}, is given by:
\begin{equation}
c_p = 4.55 - \log_{10}|T-T_{\lambda}| - 5.20\Delta,
\label{c}
\end{equation}
here $\Delta$ is a phenomenological parameter and it is assumed to be  1 at the normal-liquid phase ($t$ $>$ 0) and 0 at the superfluid phase ($t$ $<$ 0), where $t$ refers to the reduced temperature; $T$$_\lambda$ stands for the from normal-liquid-to-superfluid phase transition temperature. We stress that Eq.\,\ref{c} is in perfect agreement with experimental results \cite{Hill} in the temperature range $|$$T$$-$$T$$_{\lambda}$$|$ $=$ $|$$t$$|$ $\leq$ 200\,mK. However, in our analysis we stick to the region $|$$t$$|$ $\leq$ 10\,mK, cf.\,detailed discussion below. Yet, we stress that the expression employed for the specific heat fits was taken from an experiment carried out at constant volume. However, as the range of temperature is quite small, the process can be considerate isobaric as well as isovolumetric. Hence, in this particular case we assume $c$$_P$ $=$ $c$$_V$ so that Eq.\,\ref{Definition-GR} can be used in our analysis.
\begin{figure}[h]
\centering
\includegraphics[scale=0.3]{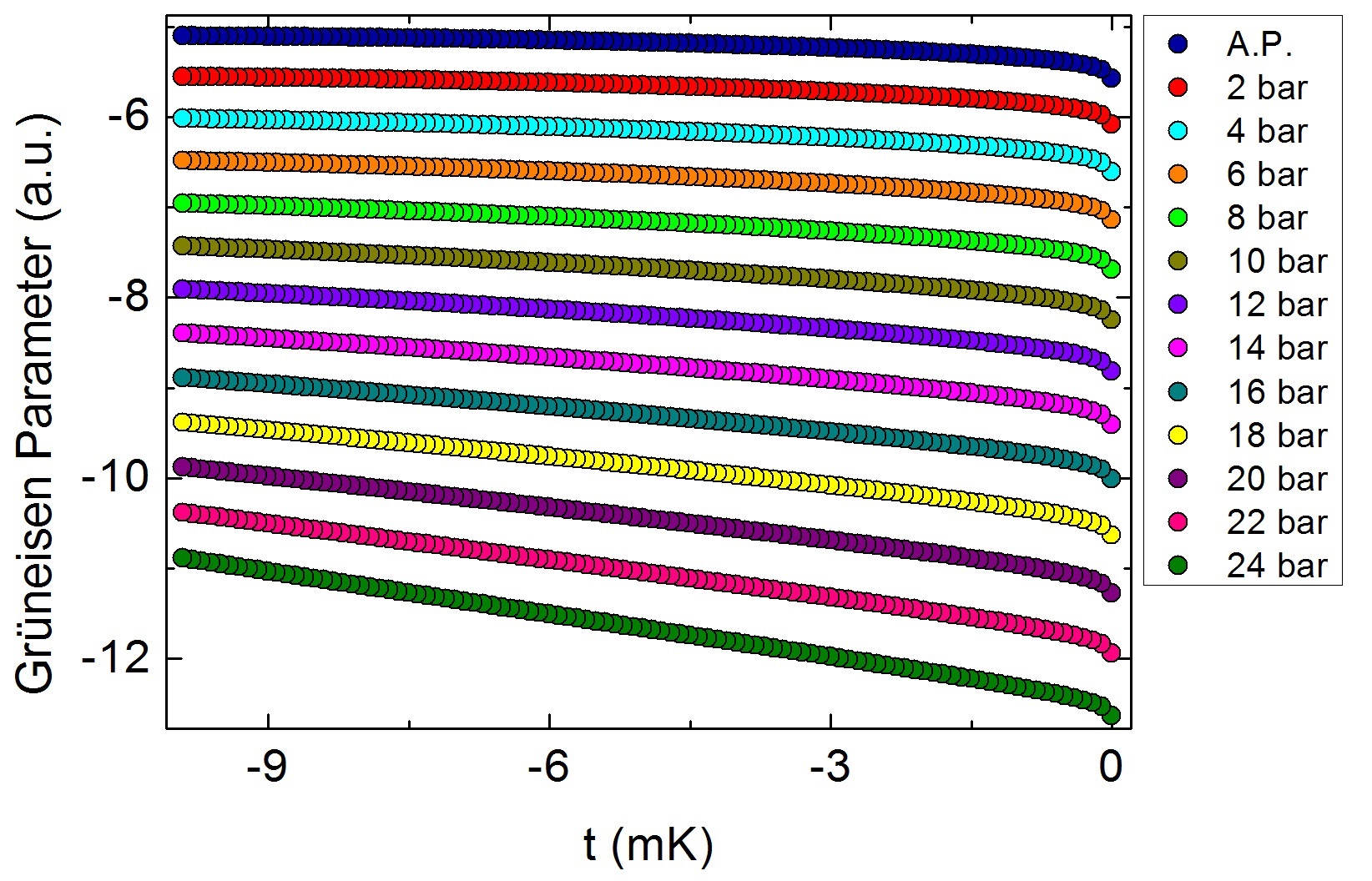}
\caption{\footnotesize Gr\"uneisen parameter as a function of the reduced temperature $t$ under various pressures, cf.\,indicated in the label. The behaviour for saturated pressure value is the same observed in Fig.\,\ref{Gruneisen} and increases with the pressure. A.P.\,stands for ambient pressure.}
\label{Gruneisen_2}
\end{figure}
Regarding thermal expansivity experiments, the work of Niemela and Donnelly \cite{Niemela} provided the data set in the $T$-range from 1.35\,K to 4.9\,K. Nevertheless, here the only results within the region $|$$t$$|$ $\leq$ 10\,mK are taken into account. The volumetric thermal expansion data are nicely described by the following expression \cite{Niemela}:

\begin{equation}
\beta(t) = -\frac{(a_1+b_1)+a_1\ln|t|+(a_2+2b_2)t+2a_2t\ln|t|}{1 + \sum\limits_{i=1}^2a_it^i\ln|t| + \sum\limits_{i=1}^7b_it^i}+
\label{Thermal}
\end{equation}
\begin{equation*}
+ \frac{\sum\limits_{n=3}^7nb_nt^{(n-1)}}{1 + \sum\limits_{i=1}^2a_it^i\ln|t| + \sum\limits_{i=1}^7b_it^i},
\end{equation*}
being,  for the sake of completeness, the coefficients $a$$_i$ and $b$$_i$ presented in Table 1.
\begin{table}[h]
\label{table1}
\centering\large
\small
\begin{tabular}{|c|c|c|c|c|}
  \hline
  $i$ & $a$$_i$ ($T$ $<$ $T$$_\lambda$) & $a$$_i$ ($T$ $>$ $T$$_\lambda$) & $b$$_i$ ($T$ $<$ $T$$_\lambda$) & $b$$_i$ ($T$ $>$ $T$$_\lambda$)\\  \hline
  1 & $-$0.00757537 & $-$0.00794605 & 0.00379937 & $-$0.0303511  \\ \hline
  2 &  0.00687483 &  0.00507051 & 0.00186557 & $-$0.0102326  \\ \hline
  3 &      -      &      -      & 0.00488345 & $-$0.00300636 \\ \hline
  4 &      -      &      -      &      -     &  0.00024072 \\ \hline
  5 &      -      &      -      &      -     & $-$0.00245749 \\ \hline
  6 &      -      &      -      &      -     &  0.00153454 \\ \hline
  7 &      -      &      -      &      -     & $-$0.00030818 \\
 \hline
\end{tabular}
\caption{{\footnotesize Parameters $a$$_i$ and $b$$_i$ used for the estimate of  thermal expansion expression of $^4$He (Eq.\,\ref{Thermal}) for $t$ $>$ 0 and $t$ $<$ 0, after Ref.\,\cite{Niemela}.}}
\end{table}
 Employing Eqs.\,\ref{c} and \ref{Thermal}, the GR as a function of temperature can be obtained, being the results in the range $|$$t$$|$ $\leq$ 0.001\,K depicted in Fig.\,\ref{Gruneisen}. Note that a clear singularity of the GR is observed in the immediate vicinity of the $\lambda$ transition for the isotope $^4$He. A blowup of the GR even closer to $T_{\lambda}$ together with the normalized Gr\"uneisen parameter is shown in the inset of Fig.\,\ref{Gruneisen}.
It is worth mentioning that for the  temperature range $|t|$ $>$ 1\,mK (not shown), at ambient pressure, the GR does not present singularities, being instead described roughly by a constant function.
 For the range $t$ $<$ 0, i.e.\,below the superfluid transition,  we were able to perform calculations of the GR as a function of temperature and pressure, whose results are presented in Fig.\,\ref{Gruneisen_2}. By doing so, we are interested now in following the pressure dependence of the entropy of $^4$He in the superfluid phase. Hence, we make use of the function for the superfluid helium entropy derived by Greywall an Ahlers \cite{greywall}, which in turn is in good agreement with experimental data published in the literature \cite{Mauer}.
\begin{figure}[h]
\centering
\includegraphics[scale=0.25]{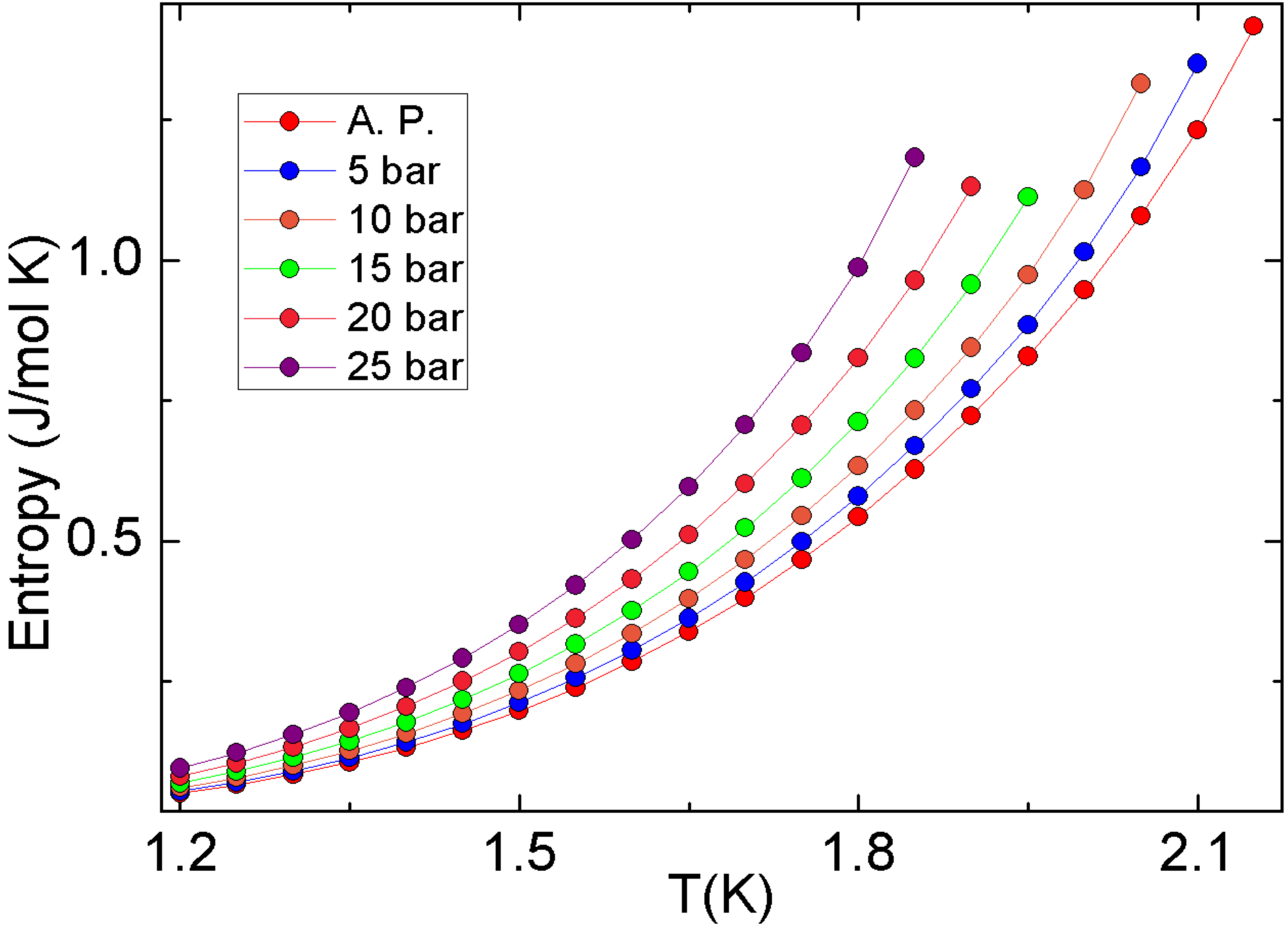}
\caption{\footnotesize Entropy as a function of temperature for various values of applied pressures \cite{Maynard}. The entropy enhancement is related to the increase of the interaction between rotons, see main text for details. A.P.\,stands stands for ambient pressure.}
\label{entropy}
\end{figure}
Interestingly enough, as can be seen in Fig.\,\ref{Gruneisen_2}, the GR is clearly enhanced under pressure. Based on the discussions presented in the introduction of this work the GR can be used as an indirect way to measure the interaction between particles in a gas, fluid or solid. In the model proposed by Tisza \cite{Tisza}, the so-called rotons are essentially a bounded pair of Helium atoms which is formed in the superfluid phase. In analogy with a classical rotor, such rotons have translational and rotational degrees of freedom.
From the sound velocity data the entropy at the superfluid phase was derived by Maynard \cite{Maynard}. Fig.\,\ref{entropy} depicts the entropy as a function of temperature for various pressures. Note that the entropy increases as the applied pressure is increased. Such a behavior is a direct consequence of the enhancement of the interaction between rotons as pressure is applied. The rotons gas is also discussed and presented in details by Pathria in Ref.\,\cite{Pathria}. Employing the expression of the Helmhotz free energy it is straightforward to obtain the GR of a rotons gas as a function of temperature and pressure (not shown), being the results in perfect agreement with those depicted in Fig\,\ref{Gruneisen_2}.

\section{Conclusions}
To summarize, we have derived the Gr\"uneisen ratio for various
gases. We have shown that for the ideal gas as well as for the free electron gas a constant value, namely $\Gamma$ = 2/3 is found for the Gr\"uneisen ratio.  For the Van der Waals gas the Gr\"uneisen ratio diverges when the gas particle co-volume achieves a critical value.
We also have shown that $\Gamma$ for $^4$He diverges near the so-called $\lambda$ transition, namely from superfluid to normal liquid. Our analysis show that application of pressure increases the system entropy, in perfect agreement with Tisza's theory based on rotons.
Using the textbook Bose-Einstein distribution for gases and
considering the boundary conditions for the realization of a
Bose-Einstein condensation we have obtained $\Gamma$ = 2/3 for the non-interacting boson gas. Yet, an analysis of the Bose-Einstein condensation in analogy with the classical liquid-to-gas transition reveals that when the gas cloud volume achieves the critical, the Gr\"uneisen ratio diverges. A calculation of the effects of interaction on the Gr\"uneisen ratio for the Bose-Einstein condensation constitute a topic of interest and will be explored in another work.

\section*{Acknowledgements}
M. de S. acknowledges financial support from the S\~ao Paulo
Research Foundation -- Fapesp (Grants No. 2011/22050-4) and National Council of
Technological and Scientific Development -- CNPq (Grants No.\,305472/2014-3).
\normalsize

\bibliography{References-Gru}

\end{document}